\def\Missing#1#2{
         \ifmmode
              {#1}\kern-0.6em\lower-.1ex\hbox{/}_{#2} 
         \else
             ${#1}\kern-0.6em\lower-.1ex\hbox{/}_{#2}$
         \fi}
\def\met{\mbox{${\hbox{$E$\kern-0.6em\lower-.1ex\hbox{/}}}_T$}} 
\def\D0{D\O}                            
\def\d0draft{}
\def\err#1#2#3 {{\it Erratum} {\bf#1},{\ #2} (19#3)}
\def\ib#1#2#3 {{\it ibid.} {\bf#1},{\ #2} (19#3)}
\def\nc#1#2#3 {Nuovo Cim. {\bf#1} ,#2(19#3)}
\def\nim#1#2#3 {Nucl. Instr. Meth. {\bf#1},{\ #2} (19#3)}
\def\np#1#2#3 {Nucl. Phys. {\bf#1},{\ #2} (19#3)}
\def\pl#1#2#3 {Phys. Lett. {\bf#1},{\ #2} (19#3)}
\def\prev#1#2#3 {Phys. Rev. {\bf#1},{\ #2} (19#3)}
\def\prl#1#2#3 {Phys. Rev. Lett. {\bf#1},{\ #2} (19#3)}
\def\rmp#1#2#3 {Rev. Mod. Phys. {\bf#1},{\ #2} (19#3)}
\def\zp#1#2#3 {Zeit. Phys. {\bf#1},{\ #2} (19#3)}

\input{epsf}
\documentclass[epj]{svjour}

\usepackage{graphics}
\begin{document}
\title{Asymptotic behavior of the Daily Increment Distribution of the IPC, the Mexican Stock Market Index}

\author{H.F. Coronel-Brizio\  \thanks{\emph{e-mail:} hcoronel@uv.mx} 
and\ A.R. Hernandez-Montoya\ \thanks{\emph{e-mail:} alhernandez@uv.mx}%
}

\institute{Facultad de F\'{\i}sica e Inteligencia Artificial.
Universidad Veracruzana, Apdo. Postal 475. Xalapa, Veracruz. M\'{e}xico.}
\date{Received: date / Revised version: date}
\abstract{
In this work, a statistical analysis of the distribution of daily fluctuations of the IPC, the Mexican 
Stock Market Index is presented. A sample of the IPC covering the 13-year period 04/19/1990 - 08/21/2003 was analyzed and the cumulative probability distribution of its daily logarithmic variations studied. Results showed that the cumulative distribution function for extreme variations, can be  described by a Pareto-Lev\'y model with shape parameters $\alpha=3.634 \pm 0.272$ and  $\alpha=3.540 \pm 0.278$ for its positive and negative tails respectively. This result is consistent with previous studies, where it  has been found that $2.5<\alpha<4$ for other financial markets worldwide.
\\
\\
\\
\\
\noindent
Presentamos un an\'alisis estad\'{\i}stico de la distribuci\'on de fluctuaciones diarias del
\'{\i}ndice de la Bolsa Mexicana de Valores, el llamado IPC (\'Indice de Precios y Cotizaciones). 
Estudiamos la funci\'on de distribuci\'on acumulativa de las diferencias logar\'{\i}tmicas diarias calculadas a partir de una muestra  del IPC que cubre un 
periodo de 13  a\~nos, que empieza el 19/04/1990 y finaliza el 21/08/2003. Hallamos que esta 
funci\'on de distribuci\'on acumulativa puede describirse para los valores extremos de estas diferencias mediante una distribuci\'on de Pareto-Lev\'y (ley potencia) con  exponentes  
$\alpha=3.634 \pm 0.272$ y  $\alpha=3.540 \pm 0.278$ en  sus colas positiva y negativa 
respectivamente. Este resultado es consistente con estudios previos que muestran que $2.5<\alpha<4$ para los  mercados financieros de diferentes partes del mundo.
\PACS{{01.75.+m}{ Science and Society - } {02.50.-r}{ Probability theory, 
stochastic processes and statistics -}{ 89.65.Gh}{ Economics; econophysics, financial markets, business and management - }{ 89.90.+n}{ Other areas of general 
interest to physicists}\vspace{-4pt}
        } 
} 
\authorrunning{Coronel-Brizio and Hernandez-Montoya.}%

\titlerunning{Asymptotic behavior of the Daily Increment distribution}

\maketitle
\section{Introduction}
\label{intro}
\vspace*{-.05cm}
\noindent
The behavior of extreme variations of economics indexes, stock prices
or even currencies, has been a topic of interest in  finance
and economics, and its study becomes relevant in the context of risk management and Financial Risk Theory. However,
these analyses are usually difficult to perform due to the small number of extreme observations in the tails of the distributions of financial time series variations. Recently, the interest of the physics community in the behavior of 
financial markets, has strongly increased, boosted by the availability of 
worldwide, electronically recorded financial data,  giving rise to a different approach
to confront  problems arising from economics. The collection of methods and
techniques originally developed in the area of physics, which are currently applied in the study
of financial complex systems, is now called Econophysics and it is becoming an emergent branch
of physics by itself~\cite{roehner,Takayasu,fisica1,fisica2,fisica3,mantegna}.\\
\noindent
In order to describe the behavior of the distribution of financial time series variations, several models have been proposed. Some of them are:

\begin{itemize}
\item Gaussian distribution~\cite{bachelier}.
\item Log-gaussian distribution (Geometric Brownian Motion)~\cite{geome}.
\item Stable Lev\'y distribution~\cite{mandelbrot,fama,mantegna1}.
\item Truncated Lev\'y Distribution~\cite{mantegna2,mantegna3,mantegna4}.
\item Poisson like distribution~\cite{baptista1,baptista2}.
\item Power law Distribution with $\alpha \simeq 3$~\cite{GOPI98} (Asymptotically).
\end{itemize}

\noindent
In this paper, a statistical analysis of the  distribution of daily  variations of the IPC\footnote{The Mexican Stock Market index or Indice de Precios y Cotizaciones for its Spanish meaning.} is presented. It is organized as follows. In the remaining of this section, we briefly review  the Pareto-Lev\'y distribution, and some of the phenomena  it describes are mentioned.  In next section, a very short introduction to the variables of common use  in the study of financial index and prices variations is given. In section 3, we introduce the sample data analyzed in this work and some important statistical properties of financial time series variations (fat tails, clustering volatility, etc)\footnote{In finance, to these statistical properties of financial time series, jointly with  other such as the properties of intermittency, asymmetry in time scales, absence of autocorrelations, gain/loss asymmetry  etc., are called ''Stylized facts'' \cite{facts}.} are discussed, all the above in the context of the IPC data. In section 4  we explain and justify the procedure to estimate the Pareto-Lev\'y exponent from  data and we show the results concerning the  fit on the tails  of the cumulative distribution of the IPC daily logarithmic variations. Finally, last section is devoted to compare our results with some other related studies of different international stock markets previously reported.
\subsection{Pareto-Lev\'y Distribution. Stable Distributions}
\vspace*{-.05cm}
\noindent
At this point, it is convenient to make a review of the definition of the Pareto-Lev\'y distribution:\\
\noindent
 An absolutely continuous random variable $Y$, is said to follow a Pareto-Lev\'y distribution with parameters $\alpha$ and $\gamma$, if
its cumulative distribution function $F$ has the form:

\begin{equation}\label{pareto1}
F(y):= P\{ Y \leq y_i\} =1- \left(\frac{y_0}{ y_i }\right)^{\alpha} = 1-\frac{\gamma}{ y_i^\alpha}
\end{equation}

\noindent
with $y_i \geq y_0$,  $y_0^\alpha = \gamma$   and  $\alpha >0$. 
When the condition $\alpha > 2$ holds, the mean and variance of Y are both finite and by the central
limit theorem, the sum of independent Pareto-Lev\'y distributed random variables, converges in probability to a gaussian law. On the other hand, when $ \alpha <2 $, the Pareto-Lev\'y distribution has infinite variance, and it is said that the distribution is stable. For a  mathematical treatment of these topics, consult~\cite{levy1} and~\cite{levy2}.  
For a review from  an econophysics point of view, references~\cite{mantegna,levy3} are recommended.\\

\noindent 
The Pareto-Lev\'y distribution is  often known as the {\em Power-law} distribution, and its 
role in Physics and other areas seems ubiquitous. In particular, we can illustrate this point  by the following examples taken from finance:  
\begin{itemize}
\item $N_{\Delta t}$, the number of trades in a given interval of time $\Delta t$, follows a power-law distribution with exponent \\
$\alpha \simeq \frac{3}{2}$~\cite{volumes1}.
\item Pareto-Lev\'y tails, with  $\alpha \simeq 3$ for extreme variations of  
individual stocks prices~\cite{plerou} and also for indexes of 
different leading stocks markets~\cite{GOPI98,GOPI99,lux}.
\item Decay of volatility correlations follows a power law distribution~\cite{model2,voldecay2}. 
\item The tail behavior of the cumulative distribution function of volatility,
is consistent with  a power law distribution with exponent $\simeq 3$~\cite{volpar}.
\end{itemize}
\noindent
All the above suggests universality in financial complex systems and in order to explain the above facts, new models and even theories, are currently being proposed~\cite{model2,model1,cubic1,cubic2}. 

\vspace*{-.05cm}
\section{Study of variations of financial time series}
\label{section:variations}
\vspace*{-.05cm}
\noindent
In the study of price variations of financial assets, many observables can be analyzed~\cite{mantegna}. 
If $Y(t)$ is the value of the index at time $t$, some commonly used observables
are:
\begin{itemize}
\item Prices or indexes changes themselves, for some interval of time $\Delta t:$  
\begin{equation}
 Z(t) := Y(t+\Delta t)- Y(t)
\end{equation}
\item Deflated prices and index changes:
\begin{equation}
 Z_D(t) : =  Z(t)\times D(t)
\end{equation}
\end{itemize}
\noindent
Where $D(t)$ is a statistical factor or index called a discount or a deflation factor, and  used to adjust the time value of money, enabling  the comparison of prices while accounting for inflation, devaluation, etc. in different time periods.
\begin{itemize}
\item Returns, defined as:
\begin{equation}
R(t) := \frac{Y(t+\Delta t)- Y(t)}{Y(t)}
\end{equation}
\item Differences of the natural logarithm of prices\footnote{Note: Some authors call returns to the 
difference of the natural logarithm of prices, and call normalized returns to our returns. For the 
case of high frequency data, each one approaches the other.},
 defined for some interval of time $\Delta t$ as:
 \begin{equation}
S(t):=\ln Y(t+\Delta t)- \ln Y(t) 
\label{diffe}
\end{equation}
\end{itemize}
\noindent
Each one having its own merits and disadvantages~\cite{mantegna}. \\
\noindent
In this analysis, we have used the former variable $S(t)$.

\section{Data sample and IPC variations}
\label{seccion:datos}
\begin{figure}[!htb]
\resizebox{0.48\textwidth}{!}{%
\includegraphics{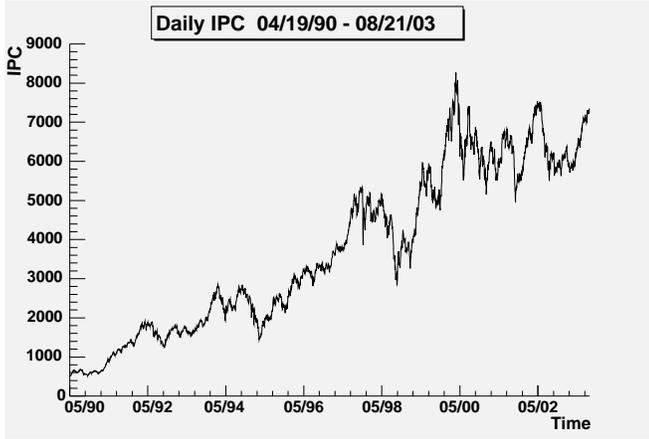}}
\caption{IPC development for the 13 year period 04/19/1990-08/21/2003.}
\label{ipc}
\end{figure}

\noindent
The database containing the IPC series analyzed here is available at~\cite{ipc_data} and covers the 13-year period 04/19/1990 - 08/21/2003. Figure \ref{ipc} shows the IPC evolution for this time period. We have used in our analysis the  daily closure values of the IPC, that is, its recorded value at the end of each trading day

\noindent
In this work, our observable is $S(t)$ as defined in equation~\ref{diffe}, and we studied the tail behavior of $P(S(t))=1-F(S(t))$, for $t=1,\ldots,N$, where $N = 3337$ is our sample size  and  $\Delta t = 1$ day.

\begin{figure}[!htb]
\resizebox{0.5\textwidth}{!}{%
\includegraphics{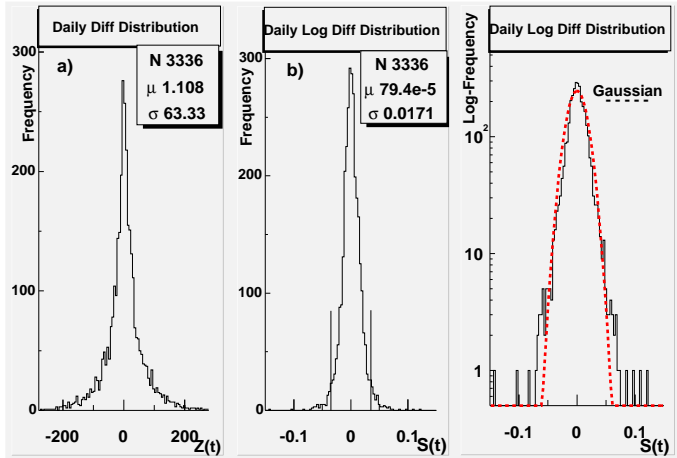}}
\caption{ IPC variations. a) Histogram of $Z(t)$, the  daily changes of the IPC index. The distribution shows symmetry and fat tails. b) Histogram of S(t). Regions studied in this paper, $S(t) > 0.035$ and $S(t)< -0.035$, are indicated by two vertical 
lines. c) Same as above, but with a vertical logarithmic scale. The broken line corresponds to a
gaussian density with the same mean and standard deviation than the $S(t)$ distribution.}
\label{diff}
\end{figure}

\noindent
Figure \ref{diff}a shows the histogram of $Z(t)$, the IPC daily changes. It is interesting that this strongly symmetric and leptokurtic (fat tailed) distribution  does not follow any well known 
model, which could appropriately describe the probability of events in its central region and in its tails at once.

\noindent
Figures \ref{diff}b and \ref{diff}c show the distribution of our observable $S(t)$. Figure \ref{diff}b shows 
that the distribution of $S(t)$ appears to follow a gaussian. This is discarded after observing figure 
\ref{diff}c,  the same distribution with a vertical logarithmic scale. There are too many extreme events visible almost as far as ten standard deviations from its mean. To easily comparing with, figure 
\ref{diff}c broken line is a gaussian scaled to the amplitude of $S(t)$ and with the same mean 
and standard deviation than those of the  $S(t)$ series.

\begin{figure}[!htb]
\resizebox{0.475\textwidth}{!}{%
\includegraphics{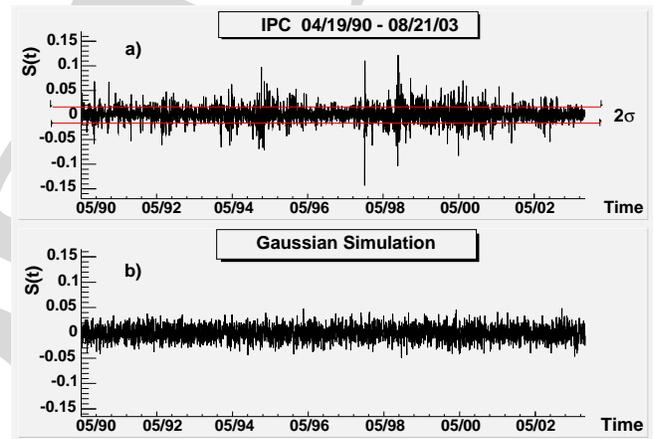}}
\caption{ a) $S(t)$ behavior for the period of time under study. Large variations in  $S(t)$, some of them as far as eight standard deviations from its mean,
can be appreciated. It can also be seen that large variations tend to form clusters in time; this
phenomena is called clustered volatility. b) Gaussian simulation already shown as a broken line in figure~\ref{diff}c. Clustering is virtually not present.}
\label{volatili}
\end{figure}

\noindent
Figure \ref{diff} shows how IPC variations are distributed; however it does not give any insight about 
the dynamics of the stochastic process that governs them. Evolution of the IPC daily 
log-differences is displayed in figure \ref{volatili}a. We observe that large variations are not 
uniformly distributed over time, and that is possible to distinguish  phases of higher volatility
\footnote{In finance, volatility is a  relative measure of price movement during a given time 
period. It can be  modeled  by the standard deviation of stock price changes. Econophysicists usually use absolute returns to model the dynamics of volatility.} alternated with phases 
of a relative financial calm. This is a characteristic of financial time series called 
clustered volatility. 

\noindent
The clustering phenomenon has not been completely understood, however some 
successful models, such as the GARCH and Stochastic Volatility models  have been proposed\footnote{The 2003 Nobel Price of Economy was awarded jointly to  Robert F. Engle for a related 
topic:''...for methods of analyzing economic time series with time-varying volatility (ARCH)''. 
Engle owns a  M. S. in Physics by Cornell University (1966).} \cite{vola1,vola2}. 

\noindent
In order to appreciate the magnitude of these strong fluctuations, gaussian values with the same  mean and standard deviation
than those of the $S(t)$ distribution were simulated.
In the simulation, nearly no clustering is appreciated, as is shown in figure \ref{volatili}b. 

\vspace*{-.05cm}
\section{Parameter estimation from IPC empirical data}
\label{seccion:analisis}
\vspace*{-.05cm}
\noindent
The procedure to estimate $\alpha$, the Pareto-Lev\'y exponent from empirical 
data is straightforward: Pareto-Lev\'y tails behave as straight lines in a 
logarithmic plot. Then, after performing a linear fit of $\log P(S(t))$ on 
$\log S(t)$, the obtained slope gives us an estimate of
the exponent $\alpha$ of the Pareto-Lev\'y Distribution. Clearly  a distribution 
whose tails do not behave linearly in a log-log plot, can not be properly 
described by the Pareto-Lev\'y  model.

\begin{figure}[!htb]
\resizebox{0.475\textwidth}{!}{%
\includegraphics{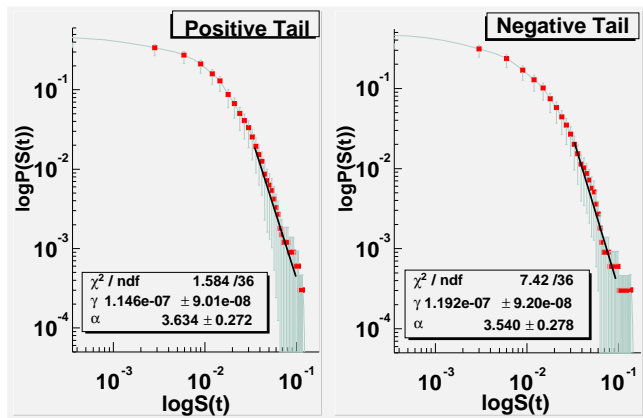}}
\caption{Linear fitted tails in a log-log plot of the cumulative distribution function $P(S(t))$ on $S(t)$. Right image positive tail. Left image negative tail. Fitted parameters are shown.}
\label{fits}
\end{figure}

\noindent
The fit was carried out using data available in the tails regions $|S(t)|>0.035$. Both tails are marked with vertical lines in figure \ref{diff}b. Those regions were chosen simply by examining the set of points for which the corresponding log-log plot behaves linearly. $P(S(t))$ was then reduced to 79 and 93 events in its negative and positive tails respectively. Note that, as is showed in figure \ref{fits}, and in order to deal with the undefined logarithmic scale for the left tail of $P(S(t))$, we have used  $-S(t)$ in our analysis.

\noindent
It was found, that for these regions, the tails of $P(S(t))$ 
decay following a power law model. A straight line provides a good fit for them in a logarithmic plot.
Both fits are shown in figure \ref{fits}.

\vspace*{-.05cm}

\noindent
 Table \ref{table1} shows  the estimated  parameters and the 95\% confidence intervals obtained from the lineal regression fit for the negative and positive tails of  $P(S(t))$.

\begin{table}[!htb]
 \begin{center}
\begin{tabular}{lccr}
\hline
\hline
Fitted region & $\alpha$ \\
\hline
\noalign{\smallskip}\hline\noalign{\smallskip}
$S(t)> 0.035$ & 3.634  $\pm$ 0.272  \\
$S(t)< -0.035$ & 3.540  $\pm$ 0.278  \\
\hline
\hline
\end{tabular}
\caption{Fitted Parameters plus minus twice the standard error of the estimates, for positive and negative tails.}
\label{table1}  
\end{center}
\end{table}

\vspace*{-.4cm}
\section{Discussion}
\vspace*{-.05cm}
\noindent
Results shown in table \ref{table1} are consistent with similar studies, where the Pareto-Lev\'y
model with $2.5 <\alpha < 4.0$, has been found  be useful for describe the behavior of extreme variations of diverse financial markets. Table \ref{table2}  summarizes results of some of these studies.

\begin{table}[!htb]
\begin{center}
\caption{Pareto exponent for some international Stock Markets. Daily data (d). High Frequency data (*).}
\label{table2}       
\begin{tabular}{lcccr}
\hline
\hline
Market & $\alpha$ (Right tail) & $\alpha$ (Left tail) & Data Period \\
\hline
\hline
AMEX & $2.84 \pm 0.12$ & $2.73\pm 0.14$ & 01/94-12/95$^*$\\
NASDAQ & & &\\
NYSE & & &\\
(Combined) & & &\\
(USA) & & &\\
~\cite{GOPI98}& & &\\

S\& P 500& $3.66\pm0.011$ & $3.61\pm 0.11$ & 1962-1996$^d$\\ 
(USA) & $3.39\pm0.05$ & $3.37\pm 0.07$ & 1984-1996$^*$\\
~\cite{GOPI99}  & & &\\

DAX& $2.4$ (minutely)& & 10/97-12/99$^{*}$\\
(Germany) &to $3.5$ (hourly) & &1959-2001$^{d}$\\
~\cite{dax,lux} & & &\\

NIKKEI & $3.05\pm0.16$ & & 1984-1997$^d$\\
(Japan) & & &\\
~\cite{GOPI99}  & & &\\

Hang-Seng & $3.03\pm 0.16$ & & 1986-1997$^d$\\
(Hong Kong) & & &\\
~\cite{GOPI99} & & &\\
\hline
\hline
\end{tabular}
\end{center}
\end{table}

\noindent
High frequency studies of price variations, most of them performed for stock markets belonging to developed countries show that the distribution of returns follows a Pareto-Lev\'y form with exponent  converging to $\alpha \simeq 3$ as $\Delta t$  decreases to time intervals of about one minute. For the case of stock markets of emergent economies, it seems they may belong to a different universality class, some studies~\cite{selcuk,mate1} show that the return distributions from emergent markets have fatter tails than the observed in developed markets.

\noindent
In summary, it has been shown that the cumulative probability of daily extreme logarithmic  changes of the Mexican IPC index, can be  approximated by the Pareto-Lev\'y model, with exponents  $\alpha = 3.634  \pm 0.272$ and $\alpha = 3.540 \pm0.278$ for its positive and negative tails respectively. As a consequence of these values, we can affirm that the stochastic process that governs the time series $S(t)$, is well outside the Lev\'y stable regime ($0 < \alpha < 2)$.\\

\textbf{Acknowledgments}

\noindent
A.R.H.M thanks Conacyt-Mexico for financial support provided under Grant No. 44598F. Also wishes to thank to N. Bagatella, N. Cruz, A. Guerra, G. Mandujano  and R.de la Mora for their useful suggestions. We specially appreciate  the bibliographical support given by P. Giubellino. \\
\noindent
Analysis in this work was performed using ROOT~\cite{root}.


\begin{thebibliography}{99}
\bibitem{roehner}
Bertrand M. Roehner, \textit{Patterns of Speculation. A Study in Observational Econophysics
} (Cambridge University Press, United Kingdom (2002) 25-35.
%
\bibitem{Takayasu}
{\textit Proceedings of the Workshop ``Empirical Science of Financial Fluctuations. The 
Advent of Econophysics''}, edited by Hideki Takayasu (Workshop Organized by Nihon Keizai 
Shimbun), Tokyo (2000).
%
\bibitem{fisica1}
J. Bouchaud, Physica A \textbf{313}, (2002) 238-251. 
%
\bibitem{fisica2}
H. E. Stanley et al, Physica A \textbf{269}, (1999) 156-159. 
%
\bibitem{fisica3}
Dietrich Stauffer, Int. J. Mod.Phys. C \textbf{11} (2000) 1081-1087.
%
%
\bibitem{mantegna}
Mantegna, R.N. and Stanley, H.E.,
{\textit An Introduction to Econophysics}, Cambridge University Press, United Kingdom, (2000).
%
\bibitem{bachelier}
L. Bachelier, Ph.D. Thesis, 
{\em Th\'eorie de la Sp\'eculation}, Annales Scientifiques de l'Ecole Normale Sup\'erieure 
\textbf{III-17}, (1900).
%
\bibitem{geome} M.F.M. Osborne, Brownian motion in the stock market. P.H. Cootner (Ed.), The Random Character of Stock Market Prices, The MIT Press, Cambridge, MA (1964) 100-128.
%
\bibitem{mandelbrot} B. B. Mandelbrot, J. Business \textbf{36} (1963) 394-419. 
%
\bibitem{fama}
E.F. Fama, J. Business \textbf{38} (1965) 34. 
%
\bibitem{mantegna1}
R. N. Mantegna, Physica A \textbf{179} (1991) 232-242.
%
\bibitem{mantegna2}
R. N. Mantegna, H.E. Stanley, Nature \textbf {376} (1995) 46.
%
\bibitem{mantegna3} R. N. Mantegna, H.E. Stanley, Phys.Rev. Lett. \textbf{73} (1994) 2946.
%
\bibitem{mantegna4} R. N. Mantegna, H.E. Stanley, J. Stat. Phys. \textbf{89} (1997) 469.
\bibitem{baptista1} M.S. Baptista, I.L. Caldas Physica A \textbf{284} (2000) 348.
%
\bibitem{baptista2} M.S. Baptista, I.L. Caldas {\it et al}, Physica A \textbf{287} (2000) 91.
%
\bibitem{GOPI98} P. Gopikrishnan  et al, Eur. Phys. J.B \textbf{3} (1998) 139-140.
%
\bibitem{facts} Rama Cont. Quant. Finance  \textbf{1} (2001) 223-236.
%
\bibitem{levy1}
P. Lev\'y, Th\'eorie de l'Addition des Variables Al\'eatories. Gauthier-Villars, Paris, (1937).
%
\bibitem{levy2}
B. V. Gnedenko and A. Kolmogorov, Limit Distribution for Sums of Independent Random Variables. Addison.Wesley, (1954) Cambridge, MA.
%
\bibitem{levy3}
J. Voit, The Statistical Mechanics of Financial Markets. Springer-Verlag. Second Edition (2003) 85-115.
%
\bibitem{volumes1}
V.Plerou, Gopikrishnan, L. A. Amaral, H. E. Stanley, Quant. Finance \textbf{1} (2001) 262-269.
%
\bibitem{plerou}
V. Plerou, P. Gopikrishnan, L. A. Amaral, M. Meyer, H. E. Stanley, Phys.Rev. E \textbf{60, 6}, (1999) 6519-6529.
%
\bibitem{GOPI99} P. Gopikrishnan et al, Phys.Rev. E \textbf{60, 5}, (1999) 5305-5316.
%
\bibitem{lux}
T. Lux, Applied Financial Economics \textbf{6}, (1996) 463-475
%
\bibitem{model2}
J. Bouchaud, Quantitative Finance \textbf{1} (2001) 105-112.
%
\bibitem{voldecay2}
P. Cizeau, Y. Liu, M. Meyer, C.-K. Peng, H. Eugene Stanley, Physica A \textbf{245} (1997) 441-445
%
\bibitem{volpar} Y. Liu, P. Gopikrishnan, P. Cizeau, M. Meyer, C. Peng, and  H. E. Stanley  Phys.Rev. E \textbf{60, 2}, (1999) 1390-1400.
%
\bibitem{model1}
S. Salomon, P. Richmond, Physica A \textbf{299} (2001) 188-197.
%
%
\bibitem{cubic1}
X. Gabaix {\it et all} Physica A, \textbf{324} (1996) 1-5.
%
\bibitem{cubic2}
X. Gabaix {\it et all}. Nature \textbf{423} (2003).
%
%
\bibitem{ipc_data} Bank of M\'exico website: http://www.banxico.org.mx
%
%
\bibitem{vola1}
A. Bera, M. Higgins. Journal of Economic Survey \textbf{7}, 305-366 (1993).
%
\bibitem{vola2}
S. Taylor, Mathematical Finance \textbf{4}, 183-204.
%
\bibitem{dax}
A.Z. G\.orski, S. Dro\.zd\.z and J. Speth, Physica A \textbf{316}, (2003) 496-510. 
%
\bibitem{selcuk}
R. Gen\c{c}ay, F. Sel\c{c}uk, B. Whitcher, An introduction to Wavelets and other filtering methods in Finance and Economics, Academic Press, San Diego 2001.
%
\bibitem{mate1}
T.D. Matteo T. Aste, M.M. Dacorogna, Physica A \textbf{324} (2003) 183-188.
%
\bibitem{root} Rene Brun and Fons Rademakers, ROOT - An Object Oriented Data 
Analysis Framework, Proceedings AIHENP'96 Workshop, Lausanne, Sep. 1996, Nucl. 
Inst.Meth. in Phys. Res. A 389 (1997) 81-86. See also http://root.cern.ch/.
%
\end{thebibliography}
\end{document}